\documentclass[fleqn,twoside]{article}
\usepackage{espcrc2}
\usepackage{multirow}

\usepackage{graphicx}
\usepackage[figuresright]{rotating}

\newcommand{\AmS}{{\protect\the\textfont2
  A\kern-.1667em\lower.5ex\hbox{M}\kern-.125emS}}

\def\Journal#1#2#3#4{{#1} {\bf #2} (#4) #3}

\def\NIMA{NIM~A}

\def\PRL{Phys.~Rev.~Lett.}
\def\PRD{Phys.~Rev.~D}

\title{ANTARES and other Neutrino Telescopes in the Northern Hemisphere.}

\author{Antoine Kouchner\address[APC]{AstroParticle and Cosmology, \\ 
       University Paris 7 Diderot, Batiment Condorcet, \\ 
       10, rue Alice Domon et Leonie Duquet, 75205, Paris cedex 13}%
   }

\begin{document}

\begin{abstract}

Several projects are concentrating their efforts on opening the high
energy neutrino window on the Universe with km-scale detectors.
 The detection principle relies on the observation, using photomultipliers, 
of the Cherenkov light emitted by charged leptons induced by neutrino interactions 
in the surrounding detector medium.\\
In the Northern hemisphere, while the pioneering {\sc Baikal} telescope, has been 
operating for 10 years, most of the activity now concentrates in the Mediterranean sea.
Recently, the {\sc Antares} collaboration has completed the construction of a 12 line array
comprising $\sim$ 900 photomultipliers.\\
In this paper we will review the main results achieved with the detectors currently in operation 
in the Northern hemisphere, as well as the R\&D efforts towards the construction of a large volume
neutrino telescope in the Mediterranean.

\vspace{1pc}
\end{abstract}

\maketitle

\section{Scientific motivations}

Neutrino astronomy has a key role to play in the present context 
towards a multi-messenger coverage of the high energy sky.
Neutrinos are indeed special probes since they can escape from the 
core of the sources and travel with the speed of light through magnetic 
fields and matter without being deflected or absorbed. 
They can therefore deliver direct information about the processes taking 
place in the production sites and reveal the existence of 
undetected sources.\\
High energy neutrinos are produced in a beam dump
scenario in dense matter via meson (mainly pion) decay, when the accelerated hadrons 
interact with ambient matter or dense photon fields: 
$$A/p + A/\gamma \rightarrow  \pi \rightarrow  \mu +  \nu_\mu \rightarrow  e +\nu_e  + 2\nu_\mu $$

As reported by the Auger experiment~\cite{auger}
, good candidates for high energy neutrino production are active galactic
nuclei (AGN) where the accretion of matter by a supermassive 
black hole may lead to relativistic ejecta~\cite{agn_jet}. Other potential sources of extra-galactic 
high energy neutrinos are transient sources like
gamma ray bursters (GRB). As many models~\cite{grb_review} for GRBs involves the collapsing of a 
star, acceleration of hadrons follows naturally. The diffuse flux of high 
energy neutrinos from GRBs~\cite{nu_grb} is lower than the one expected from 
AGNs, but the background can be dramatically reduced by 
requiring a spatial and temporal coincidence 
with the short electromagnetic bursts detected by a satellite.

High Energy activity from our Galaxy has also
been reported by ground based gamma-ray telescopes~\cite{hess}. 
Many of these galactic sources~\cite{galactic_sources} are 
candidates to accelerate hadrons and subsequently produce neutrinos. 
As explained in section~\ref{sec:detpri}, such sources 
could only be observed by a northern neutrino telescope.

\section{Detection principle}\label{sec:detpri}

The neutrino's advantage (for astronomy), the weak coupling 
to matter, is at the same time a big disadvantage (for detection).
Huge volumes need to be monitored to compensate for the feeble signal
expected from the cosmic neutrino sources. In this context, the water Cherenkov technique offers both a cheap 
and reliable option.

The detection principle relies on the observation, using a 3 dimensional array of photodetectors, of the 
Cherenkov light emitted, in a transparent medium, by charged leptons induced by charged-current neutrino 
interactions in the surrounding detector medium. 

Thanks to the large muon path length, the effective detection volume in the muon channel is substantially 
higher than for other neutrino flavors. The higher the neutrino energy the smaller the 
deviation between the muon and the neutrino
(typically $\Delta \theta \simeq \frac{0.7^{\rm o}}{(E_\nu {\rm(TeV)})^{0.6}}$), 
thus enabling to point back to the source with a precision close to the one achieved by gamma-ray telescopes. 
Muon trajectories are reconstructed using the time and amplitude from the photodetector signals.
The energy estimate (by a factor of 2-3) of the event  can be achieved thanks to the energy deposited in the detector.
Cosmic particles penetrating the atmosphere undergo a cascade of many secondary particles.
Among them, high energy muons can reach the detector and constitute 
a very intense source of background. To suppress this background the detector concentrates 
on upward detection. As a result, the field of view is restricted to 
one half of the celestial sky, below horizon. Severe quality cuts criteria are then applied to 
the reconstruction to remove remaining mis-reconstructed muons.
Atmospheric neutrinos produced in the atmospheric cascades can travel through the Earth
 and interact in the detector vicinity. To claim for an extraterrestrial discovery, one has then to search for an 
 excess of events above a certain energy (diffuse flux) or in a given direction (point sources).

\section{Current projects and detectors}\label{sec:projects}

\begin{figure}[t]
\begin{center}
\includegraphics*[width=0.5\textwidth]{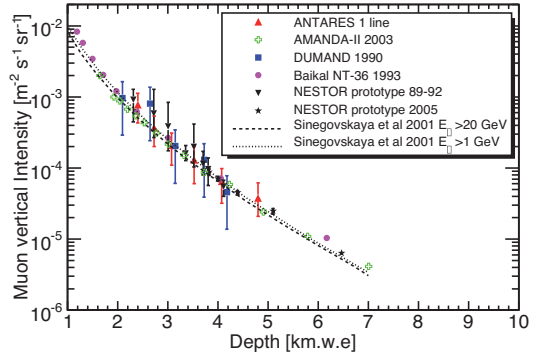}
\end{center}
\caption{Worldwild measurements of the vertical down-going muon flux, including the latest result by {\sc Antares}~\cite{antares:line1}.}
\label{fig:DIR}
\end{figure}

The first attempt to build a NT was made by the {\sc Dumand} 
collaboration~\cite{dumand}, off the Hawaiian coast. In 1987, the collaboration deployed 
a prototype string tied to a ship and performed some first measurements. 
After this success a proposal for an array of 9 strings anchored to the sea bed at 4800~m from 
the surface was submitted. Only the first line could be deployed. It was operational for several
 hours only before a leak occurred. The project was finally canceled in 1995. New projects 
were born since then, some in the ice ({\sc Amanda} and {\sc Icecube}~\cite{Icecube_isvhecri}) providing the detector with 
mechanical stability and avoiding leakage problems,
others persevering with water. Figure~\ref{fig:DIR} summarizes the muon flux measurements
of most of these projects.

\subsection{The {\sc Baikal} neutrino telescope}

\begin{figure}[t]
\begin{center}
\includegraphics[width=3in]{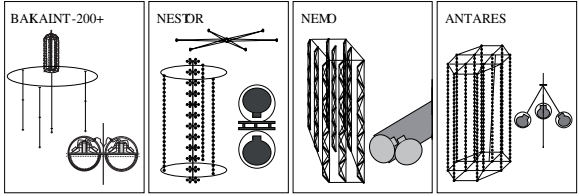}
\end{center}
\caption{From Left to right: conceptual layouts of the Baikal, Nestor and Nemo NT.}
\label{fig:geometries} 
\end{figure}

The {\sc Baikal} NT (NT-200) is located at 1100~m depth in Lake Baikal, Siberia.
Deployment and repairs are made in winter, when the lake is covered
by an ice shield.
The detector with a diameter of 40~m, consists of 192 OMs on
eight 72~m long strings (see figure~\ref{fig:geometries}).
It is running in permanent regime 
since April $6^{\rm th}$, 1998.

\begin{figure}[t]
\begin{center}
\includegraphics[height=2.3in]{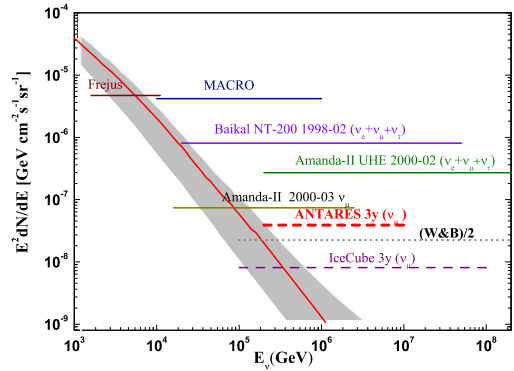}
\end{center}
\caption{Current limits (plain line), sensitivities (dash-line) and prediction (W\&B)~\cite{waxman_bound} on high energy diffuse neutrino fluxes.}
\label{fig:diffuse_limits} 
\end{figure}

The best limit on the flux of diffuse extragalactic neutrinos is obtained 
with a data sample of 1038 live days~\cite{baikal_vlvnt}(see figure~\ref{fig:diffuse_limits}).
Since directional information is not of primary importance for diffuse fluxes, 
the signal expected is an upward moving light front induced by isolated 
cascades originated by any of the 3 flavored neutrinos. To increase the sensitivity, 
no containment in the detector is 
required. The associated background is mainly bremsstrahlung showers 
initiated by high energy downward atmospheric muons. The final rejection 
criteria relies on the energy estimate of the event. A total of 372 upward going
 neutrino candidates were selected while 385 where expected by Monte Carlo simulations,
 with a median muon-angular resolution of $2.2^\circ$.

Benefiting from a low energy threshold, close to $\sim$10~ GeV (because of the high OM density), 
the {\sc Baikal} collaboration also reports interesting limits regarding WIMP as possible cold dark 
matter. Vertical neutrino events eventually produced 
by neutralino annihilations in the centre of the Earth have been looked for. In absence of excess, 
a limit on the muon flux as a function of the neutralino 
mass  has been placed close to $5\: (4) \times 10^{-15}{\rm cm}^{-2}{\rm s}^{-1}$ for a neutralino mass
of 100~GeV/${\rm c}^2$ (1~TeV/${\rm c}^2$). 

An extension of the detector called NT-200+ has been proposed and
 put into operation in April 2005. It consists of 36 additional OMs on 3 further
 140~m long strings. It is expected that a factor of 4 in sensitivity can be 
reached by this modest extension. The construction of NT200+ is a first step
 towards the deployment of a km-scale (Gton) neutrino telescope. Such a detector 
could be made of similar detection
units as NT200+, replacing the NT200 core by a single string. The detection units
 would be around 100~m distant from each other. The final detector would 
contain 1300-1700 OMs arranged on 90-100 strings with 12-16 OMs each. 
The thresholds for muons would then become 20-50~TeV while for 100~TeV 
cascades the effective volume would reach 0.5-0.8 ${\rm km}^3$. Such an extension
requires new R\&D efforts. A prototype string meant to test the future 
technology (photomultipliers, electronics, acquisition system,  trigger 
strategy, calibration elements) has been installed in April 2008 as part of the NT200+ array.

\subsection{The {\sc Nestor} project}

The {\sc Nestor} project -the first in the Mediterranean sea- started in 1989. 
The immersion site is located next to 
Pylos, on the Greek Ionian coast, at a depth of about 4000~m.  The structural 
part of the detector array is a tower rising 410~m from a sea-bed 
anchor. As shown in figure~\ref{fig:geometries}, the tower consists of twelve 6 
arm titanium 
stars of 16~m radius~\cite{nestor:design}. A pair of OMs~\cite{nestor:om} 
housing 15'' photomultipliers
in a pressure  resistant glass sphere is installed at 
the extremity of each arm. The full tower will hold 144 PMTs. The full sampled
digitized wave-form 
is sent to shore for signal analysis through a 31~km long cable deployed in 
June 2000. In January 2002 an electro-optical junction box and other devices 
dedicated to environmental survey were added. All cable connections are made 
on the boat during deployment.\\
The first {\sc Nestor} floor prototype -a reduced star (5~m radius) with 
12 OMs- has been immersed in March 2003~\cite{nestor:cern_courier}. 
More than 5 million triggers (4 fold
coincidences) have been recorded. A reconstruction has been made to extract
the zenithal dependency of atmospheric muons as well as a measurement of the
muon flux~\cite{nestor:flux}. These results validate the chosen techniques, 
even if the
collaboration still has to demonstrate its ability to deploy a full
scale rigid tower. This could be achieved with the recent construction of 
a
dedicated vessel called ``Delta-Bereniki'', made of triangular 
ballasted platform of 275 tons equipped with engines and thrusters for positioning.

\subsection{The {\sc Nemo} project}

The {\sc Nemo} collaboration was formed in 1998 and has close
ties to the {\sc Antares} experiment. The collaboration has 
located a suitable detector site, in Capo Passero, at a depth of 3500~m, 80~km 
off the eastern coast of Sicily, featuring low 
concentration of bioluminescent bacteria~\cite{nemo:site}. 
The {\sc Nemo} conceptual design of a cubic kilometer 
NT calls for 64 semi-rigid towers (which minimizes the number of in situ connections) 
. Each of
these towers would spread 16 storeys by 40~m vertical to a total length of
750~m. A sketch is also shown in figure~\ref{fig:geometries}. 
Each storey is composed of a 20~m long rigid arm that carries
two OMs at each end and hosts instrumentation for positioning and environmental parameter
monitoring. The structure is designed to be assembled and deployed in a compact
configuration. Once unfurled the floors assume an orthogonal orientation
with respect to their vertical neighbors. Communication to the shore is foreseen to be
 handled by one main and eight secondary junction boxes. \\
In order to validate the technology proposed for a km-scale detector, a Phase-1 project
was launched in 2002 and achieved, in December 2006
  with the operation, in a test site (2100~m depth,
25~km off Catania port) of a prototype mini-tower equipped with all the critical
components of a NT.  Unfortunately, after 4 months of immersion, an attenuation in the 
optical fibers transmission in the junction box was reported. A short-circuit finally occurred
in May 2007 (forcing the recovery, reparation, and re-immersion in April 2008 of the junction
box). In the mean time, a poor manufacturing process caused a loss of buoyancy to the
mini tower which slowly sank down to the sea bed. Nevertheless, a small sample of down going
 muon events was analyzed (for a total livetime of $\sim$200 hours).
Reconstructed tracks were compared to Monte Carlo simulations, revealing a good agreement in 
the shape of the zenith distributions~\cite{nemo:phase1}. \\
The {\sc Nemo} project has now entered the phase-2, that aims at operating 
a full scale tower in the Capo Passero site. The design of the tower has been modified
according to the experience gained with the phase-1 project. A 100~km long  Alcatel 
communication cable has been installed in July 2007. It links the 3500~m deep sea site to 
a brand new shore station inside the harbor area of Portopalo. The deep sea termination
of the cable, which includes a 10~kW DC/DC converter system is under realization and 
should be installed early 2009, followed by the immersion and connection of the tower, 
in spring 2009.

\subsection{First results from the {\sc Antares} NT}

{\sc Antares} is the first large scale NT constructed in the deep sea. It is located at
depth of 2475~m, $40$~km off La-Seyne-sur-Mer (Var, French Riviera) where
the shore station is installed. The deployment stage ended in May 2008.
The detector is currently running in the complete configuration. Here we 
give a short description
of the detector and focus on the results obtained with the first 5/12 of 
the detector which took data
from February to December 2007.

\subsubsection{The {\sc Antares} detector}


The final detector consists of an array of 12 flexible individual mooring lines separated 
from each other on the sea bed by 60-80~m.
The lines are weighted to the sea bed and held nearly 
vertical by syntactic-foam buoys. Each line
is equipped with 75 optical modules~\cite{antares:om}. 
 The OMs are inclined by $45^{\circ}$ with respect to the vertical axis to ensure maximum 
sensitivity to upward moving Cherenkov light fronts. A fifth of the $12^{\rm th}$ line is 
dedicated for the study of the acoustic background noise for further possible acoustic
 neutrino detection. This part is known as the AMADEUS system~\cite{antares:amadeus}. \\
The sea water properties have been extensively studied revealing low light scattering, mainly 
forward and an average optical background (induced by bacteria and $^{40}K$ decays) 
of 70 kHz per detection channel. Though the optical activity is highly variable, 
long time measurements have now confirmed these numbers~\cite{antares:neut2008}.
A mini-instrumented line equipped with 3 OMs and mainly dedicated to study environmental parameters
 has been in operation since spring 2005.
The results are presented into details in~\cite{antares:milom}.

\subsubsection{DAQ and calibration}
The default readout mode~\cite{antares:daq} of the detector is the transmission of the time and amplitude
 of any light signal above a threshold corresponding to 1/3 of a photo-electron for each OM.
 All signals are sent to shore and treated in a computer farm
 to find hit patterns corresponding to muon tracks or other physics events producing light in the water. 
Time measurements are relative to a master reference clock signal distributed from 
shore. Timing calibration is ensured by a dedicated network of laser and LED beacons.
The line motions are monitored by acoustic devices and by inclinometers
 regularly spread along the line, allowing redundancy. The system provides a location of each OM with a precision close 
to 10~cm. Together with the data from the time calibration devices, this guarantees an angular resolution 
within design expectations ($<0.5^{\circ}$) and allow to look for point source with high sensitivity, as reported
in the next section.

\subsubsection{First deep-sea neutrinos}

\begin{figure}[t]
\begin{center}
\includegraphics [height=2in]{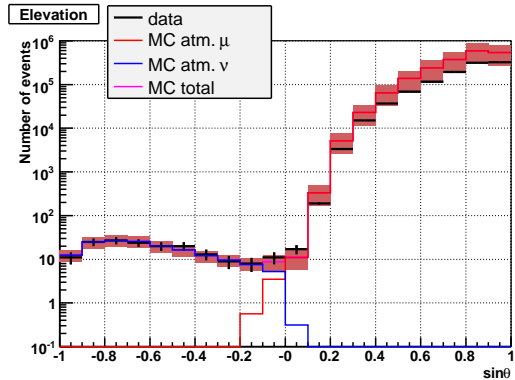}
\end{center}
\caption{The elevation angle distribution of data taken with the 5 line detector, compared to simulations.
The colored band accounts for systematic uncertainties mainly due to Monte Carlo model uncertaintis, OM acceptance and attenuation 
length.}
\label{fig:zenith}
\end{figure}

Several algorithms are available within the collaboration
for the reconstruction of a muon trajectory.
Some are specialized for reconstruction with a single line and can therefore be applied for search of vertical muons  
(figure~\ref{fig:DIR}), others can be used for search of through-going muons. A recent algorithm has been developed
which can be used on the data before the full alignment of the detector.The algorithm can therefore run quasi-online. 
The result from this algorithm is in good
 agreement with Monte Carlo simulations, as shown in figure~\ref{fig:zenith}. The sample comprises 168 neutrino 
candidates recorded over 167 active days.\\

\begin{figure}[t]
\begin{center}
\includegraphics [width=0.5\textwidth]{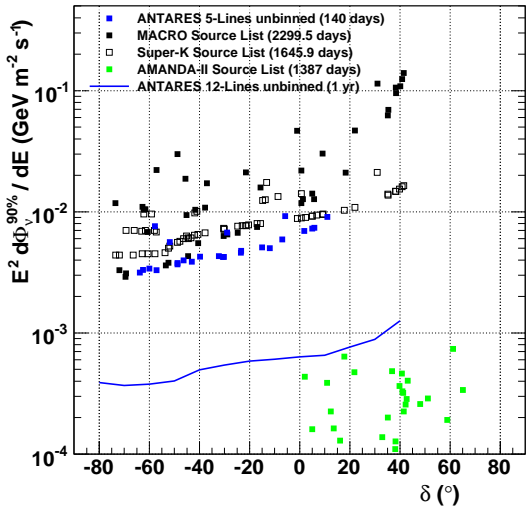}
\end{center}
\caption{Result of the search for high energy neutrinos from known point sources with the
{\sc Antares} 5-line detector. The plain line is the one year sensitivity of the full detector.}
\label{fig:pointsources}
\end{figure}

A sub-set of these data set has been used to search for steady point sources by looking for clusterization of events
on top of the atmospheric background. This required the full alignment of the detector and a more sophisticated 
reconstruction strategy  so to match the expected pointing accuracy. In order to prevent from bias, the analysis 
was first performed by scrambling the time of the events in declination bands, thus providing realistic background samples.
An unbinned method~\cite{antares:pointsource} was then applied to look for clusters in the direction of known sources.
In absence of excess, limits could be achieved as indicated on figure~\ref{fig:pointsources}. Though these limits 
have not been obtained with the final configuration of the detector, 
they already supersede the ones obtained after many years of operation by first generation experiments. 

\subsection{{\sc KM3NeT}: toward a km-scale NT in the Mediterranean.}

The three existing collaborations {\sc Antares, Nemo} and 
{\sc Nestor} have formed the {\sc KM3NeT} consortium with the goal 
to design and locate the future Mediterranean km-scale detector. Funding 
was granted by the EU in 2 different stages.\\
 First, a 3-year design phase started in February 2006, at the end of which the consortium
will deliver a
Technical Design Report defining the technological solutions for the construction.
Minimum requirements are an instrumented volume $\geq{\rm 1~km}^3$, sensitive to all
neutrino flavors and with an 
angular resolution for muon neutrinos of about $0.1^{\circ}$ in the TeV range. These 
requirements are listed into details in the ``conceptual design report''~\cite{km3net:CDR} 
published early 2008. Different possible detector configurations have 
been investigated in detailed simulation studies~\cite{km3net:dornic}.\\
 As a second stage, a preparatory phase has started in March 2008 meant to
address political, governance and financial issues related to the {\sc KM3NeT} detector, including
the site selection. The preparatory phase will also include prototyping work.
 The construction of the detector
should start in 2012.  The overall cost is estimated to be 220-250 Meuros. 
{\sc KM3NeT} is part of 
the ESFRI (European Strategic Forum on Research Infrastructures) road map
for future large scale infrastructures.
The project enlarges indeed the scope to an international and multidisciplinary endeavor.
It is 
foreseen to instrument the detector with specialized equipment for seismology,
gravimetry, radioactivity, geomagnetism, oceanography and geochemistry, making
{\sc KM3Net} a complex laboratory for a large science community.

\end{document}